# Ultra-low Energy, High Performance and Programmable Magnetic Threshold Logic

Mrigank Sharad, Deliang Fan and Kaushik Roy, *Fellow, IEEE*
School of Electrical and Computer Engineering, Purdue University, West Lafayette, Indiana 47907, USA
msharad@.purdue.edu, dfan@purdue.edu, kaushik@purdue.edu

*Abstract:* We propose magnetic threshold-logic (MTL) design based on non-volatile spin-torque switches. A threshold logic gate (TLG) performs summation of multiple inputs multiplied by a fixed set of weights and compares the sum with a threshold. MTL employs resistive states of magnetic tunnel junctions as programmable input weights, while, a low-voltage domain-wall shift based spin-torque switch is used for thresholding operation. The resulting MTL gate acts as a low-power, configurable logic unit and can be used to build fully pipelined, high-performance programmable computing blocks. Multiple stages in such a MTL design can be connected using energy-efficient ultra-low swing programmable interconnect networks based on resistive switches. Owing to memory-based compact logic and interconnect design and low-voltage, high-speed spin-torque based threshold operation, MTL can achieve more than two orders of magnitude improvement in energy-delay product as compared to look-up table based CMOS FPGA.

## 1. Introduction

Non-volatile spin-torque switches can be used in designing configurable logic blocks [1-3]. Such circuits can possibly provide enhanced scalability and energy-efficiency resulting from reduced-leakage of the spin-based memory elements. Exploiting these benefits of spin-torque devices, hybrid logic circuits like magnetic full adders [2], non-volatile flip-flops [3], and memory-cells for hybrid FPGAs have been proposed [4].

Spin devices can be attractive for logic schemes that involve direct use of memory elements for computing. One such scheme is threshold logic [5]. The operation of a threshold logic gate (TLG) involves summation of weighted inputs, followed by a threshold operation as given in eq.1:

$$Y = \text{sign}\left(\sum In_i W_i + b_i\right) \quad (1)$$

Here, $In_i$, $W_i$ and $b_i$ are the inputs, weights and the thresholds respectively. In this work we propose a magnetic threshold logic design that uses spin-torque devices for realizing input-weights as well as the thresholding switch. We employ resistive states of multi-level magnetic tunnel junctions (MTJ) for implementing configurable input weights. In such a magnetic threshold logic gate (MTLG) the inputs are received in the form of small binary voltage levels that can be less than 50mV. The resulting current-mode input received from the input resistive weights are applied to low voltage, magneto-metallic domain wall switch (DWS) [6-8]. The domain wall switch acts as a compact, low voltage and high-speed current comparator and hence provides an energy efficient threshold operation. Owing to its non-volatility, the DWS acts as a magnetic latch and hence can facilitate the design of fully pipelined logic array based on configurable MTLGs. Adjacent MTL-stages in such a design can be interconnect using resistive memory elements like CMOS compatible memristors [9-11]. The DWS based MTLGs facilitate ultra-low voltage current-mode signaling through the programmable interconnect leading to ultra-low energy dissipation in the interconnects [12, 13].

Due to afore mentioned factors MTLG can achieve high energy efficiency and performance as compared to state of the art CMOS FPGAs.

Rest of the paper is organized as follows. In section-2 we present the design of magnetic threshold logic. Performance of the proposed scheme is discussed in section 3. Section-4 concludes the paper.

## 2. Design of Magnetic Threshold logic Gate

### 2.1 Choice of design specifications

For the design of MTL-gate (MTLG) we exploit the fact that, threshold logic synthesis using gates with small fan-in (2 to 3) need fewer levels in input weights as well as reduced comparator resolution (minimum % difference between threshold and input-summation to be detected) [5]. The set of weight levels needed for different fan-in restriction is depicted in fig. 1b, which shows that for a fan-in restriction of 2, only two weight-levels are required. The number of levels in the threshold was found to be 4 in this case. Lower number of weight-levels implies higher variation tolerance for weights (fig. 1b) and relaxed resolution constraint for the comparison operation. For instance for 2-level weights, a 2-input TLG requires a comparator of only 25% resolution. The figure also shows that the increase in number of nodes while reducing the fan-in restriction from 4 to 2 is only marginal. Hence, owing to the aforementioned advantages offered by lower fan-in restrictions, in this work we limit our discussion on 2-input MTLG.

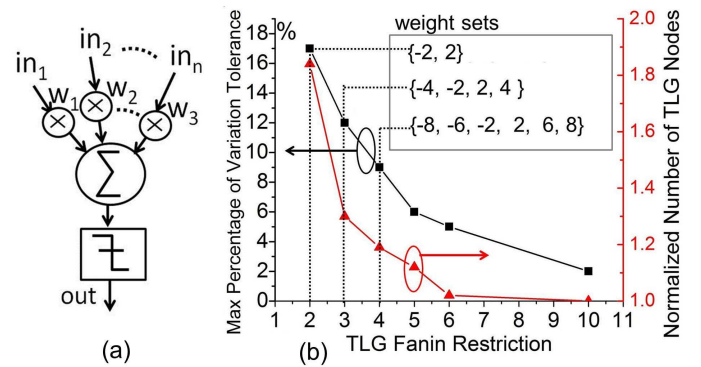

Fig. 1(a) A threshold logic gate, (b) change in number of TLG required for a given logic block for different TLG restrictions.

## 2.2 Choice of spin-devices of MTLG design

A 2-fan in MTLG requires just 2-level weights and 4-level threshold. Such weight levels can be realized using resistive memory based on magnetic tunnel junctions (MTJ) (fig 2a) [14]. The resistance of an MTJ is low (Rp) or high (Rap) depending upon parallel or anti-parallel configuration of it free and fixed magnetic layers. The ratio between Rap and Rp is determined in terms of tunnel magneto-resistance ratio (TMR) which equals (Rap-Rp)/Rp x 100. In this work a TMR of 300% has been used (i.e. Rap/Rp = 4). Input in the form of a small voltage signals are applied to MTJ-based input-weights. Current values proportional to the MTJ conductance obtained can be summed-up to realize accumulation term in the threshold-logic expression in eq.1. As depicted in fig 1b, the weights can be either positive or negative (+2 or -2 for a fan-in of 2). This can be achieved by effectively subtracting currents from 2 MTJ weights, as explained in section 2.3

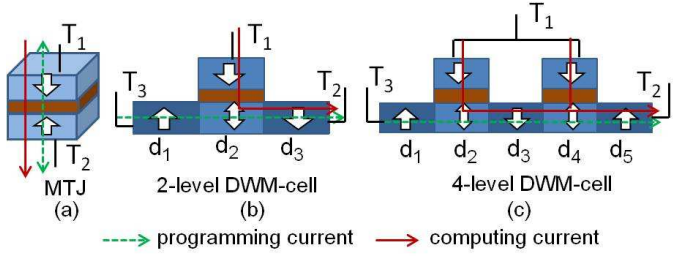

Fig. 2 (a) 2-terminal MTJ, (b) 3-terminal, two level domain wall switch, where $d_2$ is the 'free' switchable domain [14, 15] (c) three terminal, 4-level domain wall switch with $d_2$ and $d_4$ as free domains [16].

Fig. 2b shows a three terminal spin device based on domain wall magnet [14, 15]. It has a free magnetic domain $d_2$ which forms an MTJ with a fixed magnet $m_1$ at its top. The spin-polarity of $d_2$ can be written parallel or anti-parallel to the two fixed spin-domain $d_1$ and $d_3$, depending upon the direction of current flow between $d_1$ and $d_3$. The state of the free-domain can be sensed by injecting a small current between the terminals $T_1$ and $T_2$. Such a device provides decoupled read-write paths for the data-storing magnetic layer [15]. The domain wall switch shown in fig. 2b can be extended to realize to 2-bit (with resistance 4-levels) domain-wall cell shown in fig. 2c. It uses two free layers and the associated MTJs to realize 2-bit data storage and 4-resistance levels for the read path between $T_1$ and $T_2$ [16]. In this work we employ the afore mentioned domain wall switches (DWS) for implementing 2 and 4 conductance-level need for MTLG input weights and threshold respectively.

Notably, the 1-bit DWS can detect the direction or polarity (positive if going in and negative if going out of its input terminal $T_3$) of current flow across its free domain. Hence this device can be used to for current-mode thresholding operation [6-8]. The minimum magnitude of current flow required to flip the state of the free domain $d_2$ depends upon the critical current-density for magnetic domain wall motion across the free-magnetic domain $d_2$. Notably, domain-wall velocities of ~100m/s can be reached in magnetic nano-strips with current-density of the order of $10^6$ A/cm2 [17-19]. Recently application of spin-orbital coupling in the form of Spin Hall effect has been proposed for low current domain wall motion [20]. For Neel-type DW, SHE induced from an adjacent metal layer results in an effective magnetic-field ($H_{SHE}$) that can assist ST driven DW-motion (fig. 3). This phenomena can be used to reduce the switching current for the DWS device for a given switching speed [12]. In this work switching current threshold of ~2µA for 1 ns has been chosen for a DWS with free domain size of $20x2x60nm^3$. Next we describe the design of MTL gates using the DWS devices.

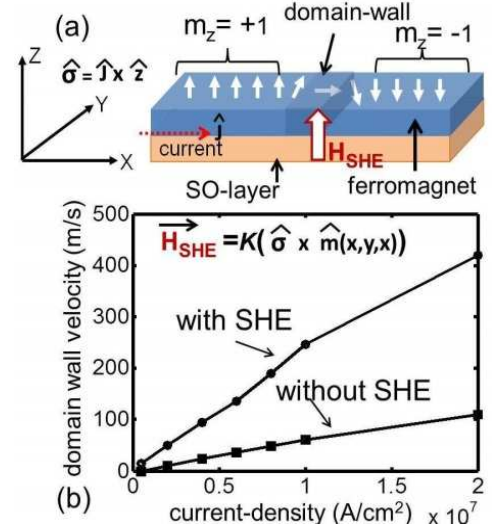

Fig. 3 (a) Domain wall magnet with SHE-coupling, (b) domain-wall velocity vs. current density, with and w/o SHE.

## 2.3 Design of MTLG using DWS devices

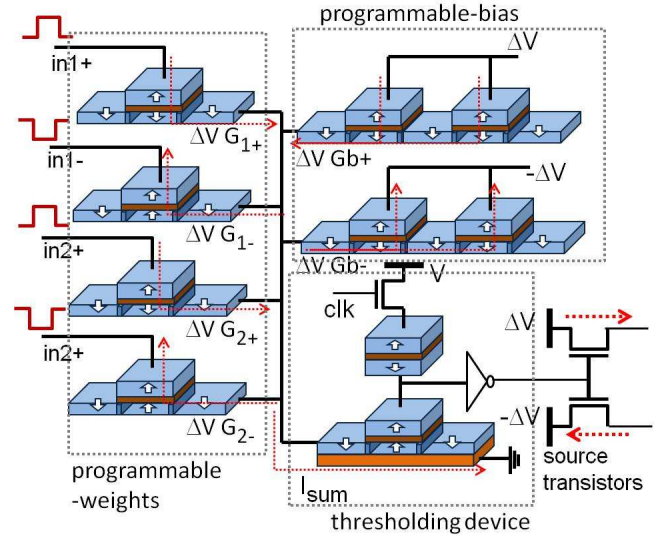

$I_{sum}= \Delta V ((G_{1+} - G_{1-})+ ( G_{2+} - G_{2-})+( G_{b+} - G_{b-})$

Fig. 4 Circuit for 2-fan in DRTL gate

Fig. 4 shows the circuit for 2-input MTL-gate based domain wall switches. The MTLG has two conductances (using DWS-MTJs) $G_{i+}$ and $G_{i-}$ for each input $in_i$. When an input is high (logic '1'), a voltage signal $+\Delta V$ and $-\Delta V$ are applied to the conductances $G_{i+}$ and $G_{i-}$ respectively, resulting in proportional current flow into the input terminal of the thresholding DWs, as shown in fig. 4. The net current due to the input $in_i$ therefore can be written as $\Delta V(G_{i+}-G_{i-})$. Thus, the two level of input weights needed for 2-fan in TLG can be obtained by programming $G_{i+}$ and $G_{i-}$ to appropriate states ( Rp or Rap). The four-levels needed for the threshold are obtained using the combination of two 2-bit DWS devices as shown in the figure. Note that two 2-bit DWS are needed to realize the bipolar 4-level weights, by connecting them to two fixed DC levels $\Delta V$ and $-\Delta V$.

The write path of the thresholding DWS is connected to ground. Using Kirchhoff's law it can be visualized that the net current flowing into the input node of the thresholding DWS is given by the following equation:

$I_{sum} = \Delta V((in_1 \cdot (G_{1+} - G_{1-}) + in_2 \cdot (G_{2+} - G_{2-}) + (G_{b+} - G_{b-}))$  (2)

This expression is essentially same as the term within the braces in eq.1. The sign function over the current-mode summation is carried out by the thresholding DWS, through the switching action of its free domain, as discussed earlier. Note that the variations in $G_i$'s can be expected to bear significant correlations due to physical proximity. The values of $G_i$'s and $\Delta V$ must be chosen such that the minimum value of $I_{sum}$ is detectable by the thresholding device. In this work the value of $\Delta V$ was chosen to be 50mV, whereas the value of Rp for the weight DWS were chosen to be ~12kΩ. this lead to the worst case input current of ~2µA for 1ns switching of the thresholding DWS.

The DWS MTJ forms a voltage divider with a reference MTJ as shown in the figure. The voltage divider is activated by the clock signal. The output voltage of the voltage divider can have a swing of Vdd/3 for a TMR of ~300%, that can be easily sensed by a CMOS inverter. The inverter in turn drives transistors that supply input currents to the fanout MTL gates.

## 2.4 Design of pipelined MTL array

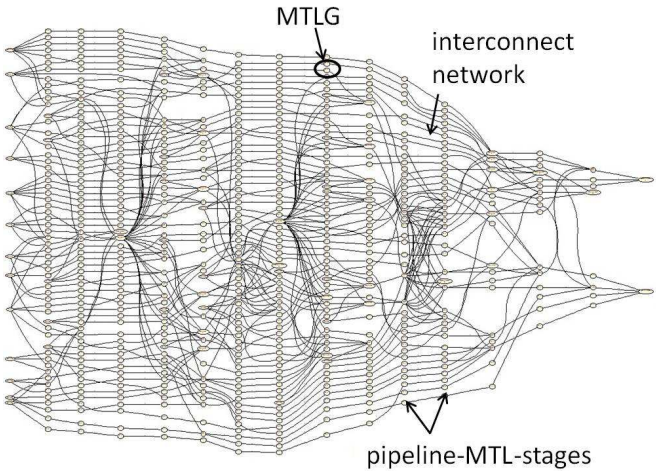

Fig. 5 Pipleined threshold logic network for the benchmark C-432

Fig. 5 shows a fully pipelined threshold logic network. It consists of TLGs arranged in multiple pipelined logic stages. In order to facilitate field programmability for the proposed MTL scheme, apart from the gate level configurability, the interconnects between the MLG levels need to be programmable. This can be achieved by the use of CMOS compatible resistive crossbar memory based on Ag-si memristors [9-11]. The range of resistance values in the resistive memory for the interconnect design is critical to the overall energy efficiency. High *ON-OFF* ratio, with low on-resistance would be desirable for energy-efficient signaling and low off-state leakage. In this work an *ON* and *OFF* resistances of ~200Ω and 1MΩ were respectively chosen. Such resistance ranges may be achievable for CMOS compatible Ag-Si memristors.

Cross-bar interconnect schematic depicted in fig. 6 shows that the gate-outputs of the $N_{th}$ MTL stage drive the corresponding metallic interconnects to $+\Delta V$ and $-\Delta V$ (for *in+* and *in-* inputs of the next stage MTLG inputs). As long as the equivalent resistance of the programmed interconnect-path is less than ~10% of the MTJ resistance of the input weights, voltage levels appearing at the input of the next stage MTLGs remain close to $\pm\Delta V$. For hardware mapping of larger logic blocks, appropriate partitioning and placement can be employed to keep the maximum interconnect length within a certain limit. In this work a maximum length of ~50µm was chosen for the interconnect, that lead to ~100Ω resistance and 10fF paracitic capacitance. It is evident that, due to ultra low voltage signalling the dynamic switching energy dissipated in MTL-interconnnect is less than 25aJ [12, 13]. This was found to be a negligibaly small fraction of the computing energy associated with the MTL gates themselves. Note that this is contrary to the present trends in conventional CMOS FPGA's where interconnects can account for upto ~90% of the total computation energy [21].

Each stage of the MTL array can be pipelied through the use of 2-phase clock for the thresholding device. The DWS being a non-voltile siwtch can fine-grained pipelining, leading to high performace and energy efficiency for a given switching current and delay for the threshold device.

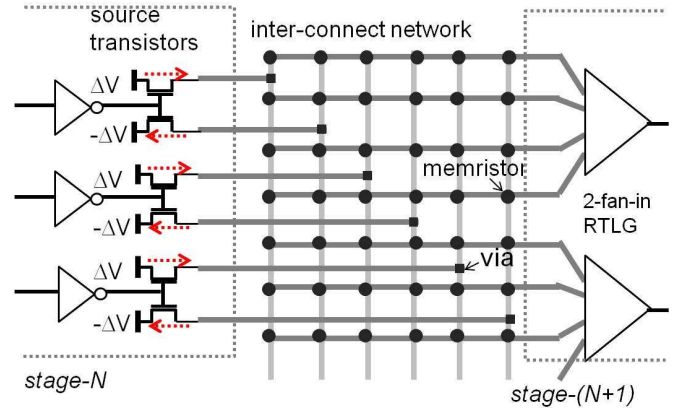

Fig. 6 Schmatic for interconnect design for DRTL using resistive crossbar memory

## 3. Design Performance

Based on the foregoing techniques for MTLG and interconnect design, we evaluated the performance of the proposed logic scheme. Table-1 compares the performance of MTL with 4-input LUT based CMOS FPGA [22], for some ISCAS-85 benchmarks. The dual functionality of computing and memory provided by the programmable spin torque switches play a significant role in energy efficiency of MTL. Ultra-low voltage operation facilitated by the DWS thresholding device reduces the static power consumption due to direct current paths. Additionally, the DWS-based MTL gates facilitate ultra-low voltage communication across the programmable resistive interconnect. The non-volatility of the DWS devices facilitate fine-grained pipelining of MTL stages, leading to high performance for a given switching delay of the thresholding device. These factors combined together lead to high energy efficiency for the proposed MTL scheme. The energy dissipation in the MTL gates depends upon the amount of static current injected per input and the applied voltage $\Delta V$. For a two input MTL with relative weight magnitudes of $[G_1, G_2, G_b]$ = [+2, +2, -3], the minimum input current injected is 50% of the current magnitude received from a single input. Thus for correct operation, the current per input should be ~2x the DWS intrinsic switching threshold. For a DWS threshold of 2μA, this implies a maximum current input of ~14μA per MTL gate. The average current input was found to be ~6μA. For $\Delta V$ =50mV and a switching time of ~1ns, this would imply an average switching energy of 0.6fJ. The power consumption in the MTJ-based voltage divider can be optimized for a desired operating frequency, by choosing higher oxide thickness. This leads to lower static current in the voltage divider and hence lower static power consumption [12]. The average power consumption in the voltage divider was found to be ~0.3μW leading to an energy dissipation of ~0.6fJ for a 2ns clock period. Thus the total energy dissipation per MTLG was close to ~1.2fJ. Average energy dynamic power dissipation in the interconnect-network was ~0.02fJ (per fan-out, based on crossbar layout), which is negligible as compared to the static power consumption in MTLG circuit. As compared to 45nm CMOS FPGA, results show ~30x lower energy more than two orders of magnitude reduction in energy-delay product, as shown in table-1.

**Table-1 Performance Comparison: MTL vs. CMOS-LUT**

| ISCAS-85 benchmark | # input | # output | delay /throughput (ns) | | Energy (fJ) | | % reduction | |
|---|---|---|---|---|---|---|---|---|
| | | | LUT | RTL | LUT | RTL | energy | energy-delay |
| c432 | 36 | 7 | 10.1 | 2 | 17362.56 | 510 | 97.1 | 99.41 |
| c499 | 41 | 32 | 8.18 | 2 | 33795.57 | 1000 | 97.04 | 99.26 |
| c880 | 60 | 26 | 8.4 | 2 | 26394.41 | 930 | 96.5 | 99.16 |
| c1355 | 41 | 32 | 9.95 | 2 | 56284.24 | 1530 | 97.28 | 99.46 |
| c1908 | 33 | 25 | 11.55 | 2 | 56930.13 | 1350 | 97.63 | 99.57 |

## 4. Conclusion

In this work we proposed the design of Magnetic Threshold Logic (MTL) that employs non-volatile spin-torque switches for reconfigurable computing. Owing to ultra-low voltage operation of the spin-torque switches and interconnects, and, high-performance, pipelined operation, MTL can achieve high energy efficiency. Comparison with 4-input LUT-based CMOS FPGA shows the possibility of ~97% higher energy efficiency and more than two orders of magnitude lower energy-delay product for MTL.

**Acknowledgment:**

This work was funded in parts by SRC and CSPIN